\documentclass[twocolumn,aps,superscriptaddress,showpacs,letterpaper]{revtex4}
\usepackage{graphicx} 
\usepackage{amsmath,amssymb}
\usepackage{dcolumn}

\newcommand{\pa}{\partial}
\newcommand{\eps}{\epsilon}

\newcommand{\DF}{\Delta F}

\newcommand{\dotx}{\dot{x}}
\newcommand{\dV}{\dot{V}}
\newcommand{\tL}{\tilde{L}}
\newcommand{\la}{\langle}
\newcommand{\ra}{\rangle}
\newcommand{\bx}{\bar{x}}
\newcommand{\bq}{\bar{q}}
\newcommand{\bS}{\bar{S}_N}

\begin{document} 
\title{Asymptotics of work distributions in non-equilibrium systems}  
\author{A. Engel}
\affiliation{Universit\"{a}t Oldenburg, Institut f\"{u}r Physik, 26111
  Oldenburg, Germany}

\begin{abstract}
The asymptotic behaviour of the work distribution in
driven non-equilibrium systems is determined using the method of
optimal fluctuations. For systems described by Langevin 
dynamics the corresponding Euler-Lagrange equation together with the
appropriate boundary conditions and an equation for the leading
pre-exponential factor are derived. The method is applied to three
representative examples and the results are used to improve the
accuracy of free energy estimates based on the application of the
Jarzynski equation. 

\end{abstract}
\date{\today}

\maketitle

\section{Introduction}
Recent progress in the statistical mechanics of non-equilibrium
systems centered around fluctuation \cite{ECM, GaCo} and work
\cite{Jar,Crooks3} theorems has profound implications for both theory and
applications. Rather complementary to the traditional emphasis of
statistical mechanics on {\it typical} behaviour of systems these new
lines of research put the {\it large deviation} properties of
thermodynamic variables like work or entropy into focus. 
Of particular interest for many practical applications is the
use of the Jarzynski equation \cite{Jar} 
\begin{equation}
  \label{eq:JE}
  e^{-\beta \DF}=\langle e^{-\beta W} \rangle
\end{equation}
to determine the free energy difference $\DF$ between two equilibrium
states at inverse temperature $\beta$ from the work distribution
$P(W)$ characterizing irreversible transitions between these
states. The method works best if $\Delta F$ is of the order of the
thermal energy, $1/\beta$. Detailed knowledge 
of free energy differences in mesoscopic systems is of crucial
importance for problems like the conformations of polymers, the
decay of metastable states, or the efficiency of molecular motors. 

It is very remarkable and particularly attractive for systems with
long relaxation times that equilibrium information like $\DF$ may be
obtained from fast changes of state. The method has been 
successfully employed in experiments on mesoscopic systems
\cite{LDSTB,DCPR,BSHSB,HSK} as well as in numerical simulations
\cite{HeJa,PaSchu} where, however, its superiority to other methods is
still under debate \cite{LeDe}. The main problems arise from the
exponential average in (\ref{eq:JE}) which is dominated by small
values of $W$ from the tail of the distribution $P(W)$. Since these
large deviations are rarely sampled the resulting free
energy estimate may be poor. An equivalent observation is
\cite{Ritort,Jar06} that the {\it dominant} trajectories contributing
most to the average in the r.h.s of (\ref{eq:JE}) are in general rather
different from the {\it typical} ones, i.e. from those with the
highest probability. Several methods have been put forward to improve
the accuracy of free energy estimates by, e.g., including information
from the backward process \cite{Bennett,Crooks3,Shirts,Collin}, using
mappings and auxiliary drifts \cite{Jar2002,VaJa,HaTh}, or
implementing biased path ensembles \cite{Sun,YtZu,ObDe}. 

In the present paper we devise a method to analytically determine the
asymptotics of the work distribution $P(W)$ of driven Langevin systems
for very small or large values of $W$. We demonstrate that fitting
these asymptotics to the region of work values that is still
sufficiently sampled by experiment or simulation significantly
improved estimates of the free energy difference may be obtained. 

The procedure builds on the method of optimal fluctuation which 
rests on the general assumption of large deviation theory 
\cite{WeFr,Touchette} that the probability of an unlikely event is
dominated by the most probable fluctuation giving rise to it. All
other possibilities to bring the same result about are even more
unlikely and may be safely neglected. In physical context the method
was originally proposed to determine the asymptotic tail of the
electronic density of states in random potentials
\cite{Lifshitz,HaLa,ZiLa}. Later applications include the motion of
charge density waves in disordered media \cite{Fei}, the velocity
distribution in Burgers turbulence \cite{FKLM,BFKL}, anomalous optical
absorption in disordered semiconductors \cite{MAK}, and the free
energy distribution of a directed polymer in a random medium
\cite{KoKo}. Recently there have also been applications to optimal
control theory \cite{VuRa} and error correcting codes \cite{CCSV}. In
the present example of Langevin dynamics the method corresponds to a
saddle-point approximation in a functional integral over stochastic
trajectories.  

The paper is organized as follows. In section 2 we outline the general
theory. First the optimal path for a work value in the tail of $P(W)$
is determined by the solution of a variational problem, then the
contribution from neighbouring paths is included. Section 3 concerns the
discussion of three concrete examples. For the first the complete
$P(W)$ can be determined analytically so it serves merely as a test of
our method. In the second we compare our results with numerical
simulations of the Langevin equation whereas the third uses
experimental data. Finally, section 4 contains some conclusions.

\section{General Theory}
For concreteness we consider a system with overdamped Langevin
dynamics described by  
\begin{equation}
  \label{eq:LE}
  \dot{x}=-V'(x,t) +\sqrt{2/\beta}\; \xi(t)\; ,
\end{equation}
where $x$ denotes the degrees of freedom, $V$ is a potential giving
rise to a deterministic drift, and $\xi(t)$ is a standard Gaussian white 
noise source obeying $\la \xi(t)\ra\equiv 0$ and $\la \xi(t)\xi(t')\ra=
\delta(t-t')$. We denote derivatives with respect to $x$ by a prime
and those with respect to $t$ by a dot.

During the time interval $[0,t_1]$ the potential  
changes between $V_0(x)=V(x,0)$ and $V_1(x)=V(x,t_1)$ according to a
fixed protocol. Using prepoint discretization the probability density
functional of trajectories starting at $t=0$ at $x_0$ and ending at
$t=t_1$ at $x_1$ is up to a constant given by 
\begin{equation}
  \label{eq:defpofx}
  p[x(\cdot)|x_0,x_1]\sim\exp\Big(
      -\beta\int_0^{t_1} \!\! dt \; L(x(t),\dotx(t),t)\Big)\; ,
\end{equation}
with the Lagrangian
\begin{equation}
  \label{eq:defL}
  L(x,\dotx,t)=\frac{1}{4}\Big(\dotx+V'(x,t)\Big)^2\; .
\end{equation}
The initial point, $x_0$, is sampled from the Gibbs measure
corresponding to $V_0(x)$ whereas the final point, $x_1$, is free. 
For the work performed along a particular trajectory $x(t)$ we have
\cite{Sekimoto} 
\begin{equation}
  \label{eq:defwork}
  W[x(\cdot)]=\int_0^{t_1} \!\!dt\; \dot{V}(x(t),t)\; .
\end{equation}
With the initial partition function 
\begin{equation}
  \label{eq:defZ}
  Z_0=\int dx \; e^{-\beta V_0(x)}
\end{equation}
the probability distribution of the work is given by 
\begin{widetext}
\begin{equation}\label{eq:defPofW}
  P(W)=\frac{1}{Z_0}\int dx_0  \exp(-\beta V_0(x_0)) \int dx_1
       \int\limits_{x(0)=x_0}^{x(t_1)=x_1} {\cal D}x(\cdot)\, 
       p[x(\cdot)|x_0,x_1]\;\delta(W- W[x(\cdot)])
\end{equation} 
Using (\ref{eq:defpofx}),(\ref{eq:defL}), and (\ref{eq:defwork})
we then find
\begin{equation}\label{eq:PofW}
  P(W)=\int \frac{dx_0}{Z_0}\int dx_1 \int\frac{dq}{4\pi/\beta} 
     \int\limits_{x(0)=x_0}^{x(t_1)=x_1} {\cal D}x(\cdot)\;
        e^{-\beta S[x(\cdot),q]}   
\end{equation}
with the action 
\begin{equation}
  \label{eq:defS}
 S[x(\cdot),q]= V_0(x_0)+\frac{1}{2}\int\limits_0^{t_1} dt\;
   [\frac{1}{2}(\dot{x}+V')^2+iq\dot{V}]-\frac{i}{2}qW\, .
\end{equation}
\end{widetext}
To apply the method of optimal fluctuations in the present context we
evaluate the integrals in (\ref{eq:PofW}) for a prescribed value of $W$
by the saddle-point approximation. Formally this corresponds to
considering the weak noise limit $\beta\to\infty$. 

\subsection{The optimal trajectory}
The determination of the optimal trajectory in (\ref{eq:PofW})
includes the optimal choice of its initial and final
point \cite{Sascha}. Introducing the augmented Lagrangian 
\begin{equation}
  \label{eq:defLt}
  \tL(x,\dotx,t)= L(x,\dotx,t)+i\frac{q}{2} \dot{V}(x(t),t)
\end{equation}
the corresponding Euler-Lagrange equation (ELE) takes the form
\begin{equation}\label{eq:ELE}
  \frac{d}{dt} \frac{\pa\tL}{\pa \dotx}-\frac{\pa \tL}{\pa x}=0\; .
\end{equation}
It is completed by the natural boundary condition 
\begin{equation}
  \frac{\pa\tL}{\pa \dotx}\Big|_{t=t_1} =0 \label{eq:bc1}\; ,
\end{equation}
at the end of the interval and the initial condition
\begin{equation}\label{eq:bc0}
  \frac{\pa\tL}{\pa \dotx}\Big |_{t=0} - V_0'(x_0)=0\; ,
\end{equation}
incorporating the sampling of the starting point from the equilibrium
distribution at $t=0$. Solving (\ref{eq:ELE})-(\ref{eq:bc0}) and
eliminating the Lagrange parameter $q$ using (\ref{eq:defwork}) we
generically find for each value of $W$ exactly one optimal trajectory
$\bx(t;W)$. The asymptotic estimate 
\begin{equation}
  \label{eq:asym1}
  P(W)\sim e^{-\beta S[\bx(\cdot),\bq]}
\end{equation}
for the distribution of work values becomes the more accurate the
larger $\beta$ is or, equivalently, the more $W$ lies in the tail of 
$P(W)$. 

\subsection{Neighbourhood of the optimal trajectory} 
Although (\ref{eq:asym1}) gives a correct estimate of the asymptotic
behaviour of $P(W)$ it is often desireable to improve its accuracy by
incorporating the dominant pre-exponential factor. This factor
has contributions from trajectories in the neighbourhood of the 
optimal one and also accounts for the Jacobian accompanying the transition
from $p[x(\cdot)]$ to $P(W)$. It is determined by including the
quadratic fluctuations around the saddle-point into the
calculation. This can be accomplished by adopting the Gelfand-Yaglom  
method \cite{Montroll,GeYa,ChDe} which yields an ordinary differential
equation for the fluctuation determinant to the present problem.

Two points are different from the standard case. 
First, the free endpoints of the optimal trajectory contribute to
the Gaussian fluctuations and give rise to modified boundary 
conditions for the fluctuation determinant. Second, the constraint
$W[x(\cdot)]=W$ suppresses some fluctuations and gives rise to a
correction factor to the free fluctuation determinant. Some details of
the explicit calculation necessary to incorporate these two
modifications are given in the appendix. 

Using $\bar{S}=S[\bx(\cdot),\bq],\, \bar{V}=V(\bx(t),t)$ and similarly
for derivatives of $V$ the final result for the asymptotics of the
work distribution is 
\begin{equation}
  \label{eq:asym}
  P(W)=\frac{e^{-\beta\bar{S}}}{Z_0 \sqrt{R\; Q(t_1)}}
     (1+{\cal O}(1/\beta))
\end{equation}
where $Q(t)$ is the solution of the initial value problem 
\begin{align}\nonumber
  0&=\ddot{Q}+2\bar{V}''\dot{Q}+
   [(2-i\bq)\dot{\bar{V}}''+(\dot{\bx}-\bar{V}')\bar{V}''']Q\\
  Q&(t=0)=1\qquad \dot{Q}(t=0)=0 \label{eq:resQ}
\end{align} 
and $R$ is given by
\begin{equation}
  \label{eq:defR}
  R=\int_0^{t_1}\!\!dt \int_0^{t_1}\!\!dt'\, \dV'(\bx(t),t)
   \Big[\frac{\delta^2 \bar{S}}{\delta x(t) \delta x(t')}\Big]^{-1}
    \dV'(\bx(t'),t')\, .
\end{equation}

\section{Examples}

\subsection{The shifted parabola}

As a first example we consider a Brownian particle dragged in a 
parabolic potential, i.e.
\begin{equation}
  \label{eq:defV1}
   V(x,t)=(x-t)^2/2\, .
\end{equation}
This system has been been
analyzed thoroughly both from the theoretical \cite{MaJa,ZoCo,Cohen}
as well as from the experimental side \cite{Wang+}. The distribution
$P(W)$ is known to be Gaussian 
\cite{MaJa,ZoCo} 
\begin{equation}\label{eq:Pexact}
  P(W)=\sqrt{\frac{\beta}{2\pi\sigma_W^2}}
     \exp\Big(-\beta\,\frac{(W-\sigma_W^2/2)^2}{2\sigma_W^2}\Big) 
\end{equation}
with 
\begin{equation}
  \sigma_W^2=2(t_1-1+e^{-t_1})\; .
\end{equation}
Since in this example the complete distribution $P(W)$ is known
exactly it merely serves as a test of our method. 

The ELE (\ref{eq:ELE}) is for (\ref{eq:defV1}) linear and can be
solved analytically with the result 
\begin{equation}\nonumber
  \bx(t;W)=\frac1 2 (2t+e^{-t}-e^{t-t_1})
           -\frac{W(2-e^{-t}-e^{t-t_1})}{2(t_1+e^{-t_1}-1)}\, .
\end{equation}
This yields  
\begin{equation}
  S[\bx(\cdot),\bq]=\frac{(W-(t_1+e^{-t_1}-1))^2}
         {4(t_1+e^{-t_1}-1)} \label{eq:hores}
\end{equation}
which correctly reproduces the exponential factor in
(\ref{eq:Pexact}). The explicit form  $\bx(t;W)$ of the optimal
trajectory for different values of $W$ and $t_1$ characterizes the
optimal combination of unlikely initial condition $x_0$ and rare
realization of the noise $\xi(t)$ necessary to bring about large
deviations in $W$.

To determine the prefactor in (\ref{eq:Pexact}) we first observe that
for (\ref{eq:defV1}) the differential equation (\ref{eq:resQ}) reduces
to  
\begin{equation}
  \ddot{Q} +2\dot{Q}=0,\quad Q(0)=1,\; \dot{Q}(0)=0
\end{equation}
with the solution $Q(t)\equiv 1$. Moreover
\begin{equation}
  \frac{\delta^2 \bS}{\delta x(t)\delta x(t')}=
     -\frac{1}{2}\,\delta''(t-t')+\frac{1}{2}\,\delta(t-t')\, .
\end{equation}
Combining this expression with the boundary conditions 
$\dot{y}(0)=y(0)$ and $\dot{y}(t_1)=-y(t_1)$ for the fluctuations
around the optimal path yields 
\begin{equation}
  \left[\frac{\delta^2 \bS}{\delta x(t)\delta x(t')}\right]^{-1}
   =-2\theta(t-t')\sinh(t-t')\,+\,e^{t-t'}\, .  
\end{equation}
With $\dot{V}'\equiv-1$ we then find 
\begin{equation}
  R=2(t_1-1+e^{-t_1})=\sigma_W^2 \, .
\end{equation}
Putting all together and using $Z_0=\sqrt{2\pi/\beta}$ the prefactor
of (\ref{eq:Pexact}) is also reproduced. In this simple example the
asymptotic result hence already gives the complete distribution.

\subsection{The breathing parabola}

A more advanced example \cite{Sascha} is given by the breathing
parabola \cite{Jar3,Carberry+}, 
\begin{equation}\label{eq:brepa}
  V(x,t)=\frac{k(t)}{2}\,x^2
\end{equation}
for which the distribution of work is neither Gaussian nor completely
accessible analytically. We will consider the case of a monotonously
decreasing function $k(t)$ implying $W\leq 0$ and determine the
asymptotic form of $P(W)$ for $W\to-\infty$. The ELE (\ref{eq:ELE}) is
given by  
\begin{equation}
  \label{eq:ELE2}
  \ddot{x}+\big((1-iq)\dot{k}-k^2\big)x=0
\end{equation}
whereas the boundary conditions (\ref{eq:bc0}) and (\ref{eq:bc1})
acquire the form
\begin{equation}
  \dot{x}(0)=k(0) x_0 \qquad\text{and}\qquad \dot{x}(t_1)=-k(t_1)x_1
\end{equation}
respectively. These equations constitute a Sturm-Liouville eigenvalue
problem which for the special choice 
\begin{equation}
  \label{eq:defk}
  k(t)=\frac{1}{1+t}
\end{equation}
can be solved analytically. The result is 
\begin{align}\nonumber
   \bx(t;W)=& \pm\frac{\sqrt{-W}}{\sqrt{g(\mu)}}\sqrt{1+t}\\
            & \Big(2\mu\cos(\mu\ln(1+t))+\sin(\mu\ln(1+t))\Big)
   \label{eq:resELE}
\end{align}
where 
\begin{equation}\label{eq:resg}
   g(\mu)=\frac{1}{2}\Big[(\mu-\frac{1}{4\mu})\sin\nu-\cos\nu
          +1+\nu(\mu+\frac{1}{4\mu})\Big]>0.
\end{equation}
and $\mu=\sqrt{iq-9/4}$ is a solution of  
\begin{equation}
  \label{eq:EV}
  (4\mu^2-3)\sin\frac{\nu}{2}-8\mu\cos\frac{\nu}{2}=0\,
\end{equation}
with $\nu=2\mu\ln(1+t_1)$.

\begin{figure}[t]
\begin{center}
\includegraphics[width=0.45\textwidth]{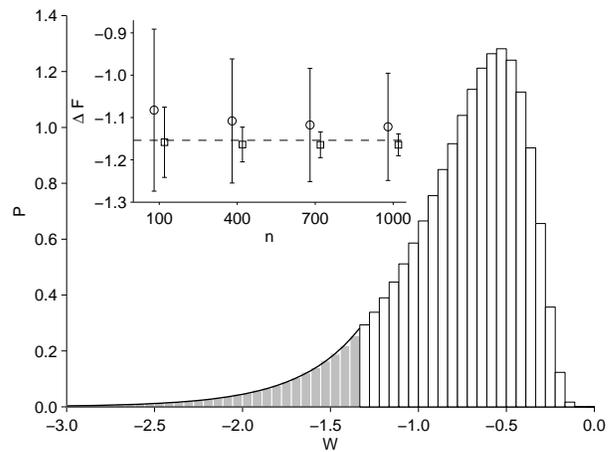}
\end{center}
\caption{\label{fig:hist} Histogram of work values as obtained from  
  $10^7$ simulations of the Langevin equation (\ref{eq:LE}) with 
  $\beta=2$, $t_1=100$, and $V(x,t)$ given by (\ref{eq:brepa}) and
  (\ref{eq:defk}). The full line is the asymptotic form of the work
  distribution as derived in (\ref{eq:res2}). The inset shows 
  two estimates for the free energy difference $\Delta F$ together
  with their standard deviation as function of the sample size
  $n$. Circles are the standard estimate (\ref{eq:Jest}), squares
  give the improved one (\ref{eq:impest}) incorporating the asymptotic
  behaviour of $P(W)$ with $W^*=-4/3$. The dashed line is the
  exact result.}  
\end{figure} 

There are hence infinitely many discrete values 
$\bq_0, \bq_1, ...$ of $q$ each associated with two trajectories
$\bx^{n+}(t;W)$ and $\bx^{n-}(t;W)$ related to each other by the
inversion symmetry $x\to -x$ of the problem. All $\bx^{n\pm}(t;W)$ are
local maxima of $p[x(\cdot)]$. However, it can be proved that  
$p[\bx^{0\pm}(\cdot)]>p[\bx^{n\pm}(\cdot)]$ for all $n>0$, i.e. the
maxima at $\bx^{0\pm}(t;W)$ are the dominant ones. This is in accordance
with intuition since large absolute values of $W$ are realized
by trajectories which are most of the time far from the minimum of the
potential. On the other hand it is known from the general theory of
Sturm-Liouville problems that the $\bx^{n\pm}(t;W)$ have $n$ zeros in
the interval $(0,t_1)$. It is hence not surprising that the ``ground
state'' solutions $\bx^{0\pm}(t;W)$ dominate the asymptotics of
$P(W)$. 

Neglecting contributions from the sub-dominant maxima we hence find
from (\ref{eq:resELE}), (\ref{eq:brepa}), and (\ref{eq:defS}) 
for the exponential term in the asymptotic work distribution  
\begin{equation}
  \label{eq:res2a}
   P(W)\sim e^{-\beta S[\bx^{0\pm}(\cdot),\bq_0]}=e^{\beta h(\mu_0)W} 
\end{equation}
where 
\begin{align}\nonumber
  h(\mu)=&\frac{1}{8g(\mu)}\Big[(8\mu^2-6)\cos\nu-
        (2\mu^3-11\mu+\frac{9}{8\mu})\sin\nu\\
         &+8\mu^2+6+\nu(2\mu^3+5\mu+\frac{9}{8\mu})\Big]\, .
\end{align}
Using (\ref{eq:resg}) one can show that $h(\mu)>1$
as is necessary for the existence of the Jarzynski
average~(\ref{eq:JE}). Note also that in the present case different
values of $W$ do not correspond to different values of $\bq$ since the
latter is fixed. As shown by (\ref{eq:resELE}) different values of $W$
are realized by different initial conditions of $\bx^{0\pm}(t;W)$. 

In the determination of the pre-exponential factor to the asymptotic
we concentrate on its dependence on $W$. From (\ref{eq:resQ}) we find
using 
(\ref{eq:brepa}) 
\begin{equation}
  0=\ddot{Q}+2 k(t)\dot{Q}+(2-i\bq_0)\dot{k}(t) Q
\end{equation}
and therefore $Q(t_1)$ is independent of $W$. Likewise 
\begin{equation}
  \frac{\delta^2 \bS}{\delta x(t)\delta x(t')}=
     -\frac{1}{2}\,\delta''(t-t')+\frac{1}{2}(k^2-(1-\bq_0)\dot{k})
     \delta(t-t')
\end{equation}
does not depend on $W$ and hence neither does its inverse. On the
other hand $\dot{\bar{V}}'=\dot{k}\bx^{0\pm}$ is proportional to
$\sqrt{-W}$ as follows from (\ref{eq:resELE}). This implies
$R\sim\sqrt{-W}$ and we get the asymptotic result
\begin{equation}
  \label{eq:res2}
  P(W)\sim\frac{e^{\beta h(\mu_0) W}}{\sqrt{-W}}\; .
\end{equation}

\begin{figure}[t]
\begin{center}
\includegraphics[width=0.45\textwidth]{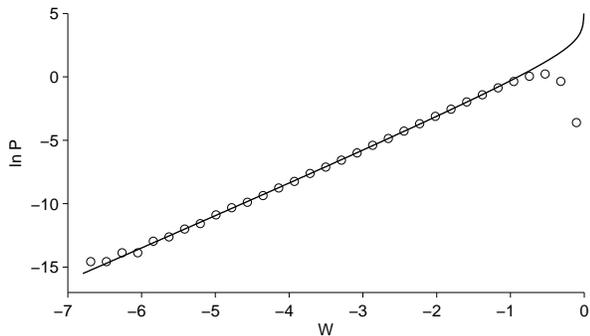}
\end{center}
\caption{\label{fig:asylog} Comparison of the simulation results
  (circles) for $P(W)$ shown in Fig.~\ref{fig:hist} and the asymptotic
  behaviour (\ref{eq:res2}) (full line) on a logarithmic scale.}
\end{figure}

It is instructive to check this result against numerical simulations of
the Langevin dynamics \cite{Sascha}. Fig.~\ref{fig:hist} shows a
histogram of work 
values obtained from such simulations together with the asymptotics
(\ref{eq:res2}). Fig.~\ref{fig:asylog} provides a logarithmic blowup
of the small-$W$ region. To determine the prefactor in 
(\ref{eq:res2}) a breakpoint $W^*$ is chosen and the area under
the asymptotic form of $P(W)$ for $W<W^*$ is set equal to
the total weight $\Phi_<$ of the histogram for $W<W^*$ (grey bars in
Fig.\ref{fig:hist}). The value of $W^*$ has to be chosen such
that on the one hand $P(W)$ is already well approximated by its
asymptotic form (\ref{eq:res2}) and on the other hand the region
around $W^*$ is still sufficiently sampled by the histogram. As shown
by Fig.~\ref{fig:asylog} in the present case there is a whole window
of admissible values of $W^*$ extending from roughly -1.5 down to
around -5. 

The asymptotic form of $P(W)$ can be utilized to improve the estimate
(\ref{eq:JE}) for the free energy difference $\Delta F$. To show this
we have subdivided the $10^7$ work values $W_i$ obtained in the
simulations into $10^4$ runs. Using $n=10^2 ...10^3$ values from each
run we have then determined the standard Jarzynski estimate 
\begin{equation}
  \label{eq:Jest}
  \Delta F^{\mathrm{st}} =
  -\frac{1}{\beta}\ln\left(\frac{1}{n}\sum_{i=1}^n e^{-\beta
      W_i}\right)  
\end{equation}
as well as an improved one 
\begin{equation}\label{eq:impest} 
 \Delta F^{\mathrm{im}} = -\frac{1}{\beta}\ln
 \Big(c\int\limits_{-\infty}^{W^*}
   \frac{e^{(h-1)\beta W}}{\sqrt{-W}} dW 
    +\frac{1}{n}\!\!\!\sum_{W_i\geq W^*}\!\!\! 
    e^{-\beta W_i}\Big)
\end{equation}
using the asymptotic form of $P(W)$ for $W<W^*$. Here the constant $c$
is determined from the normalization condition 
\begin{equation}
  \label{eq:quant}
  \Phi_<=c\int\limits_{-\infty}^{W^*}\frac{e^{h\beta
      W}}{\sqrt{-W}}\,dW\, .
\end{equation}
The inset in Fig.\ref{fig:hist} shows both estimates together with
their standard deviation for different values of $n$ as well as the
exact result $\Delta F^{\mathrm{exact}}=-\ln(1+t_1)/(2\beta)$. As is
clearly seen both the bias and the standard deviation are
significantly reduced when combining the histogram with the asymptotic
form of $P(W)$ as given by (\ref{eq:res2}).

\subsection{Driven Brownian particle near a wall}

We finally demonstrate the applicability of our method to the analysis
of experimental data. In \cite{BSHSB} a charged colloidal
particle near a wall was subjected to a time-dependent anharmonic
potential $V(x,t)$ generated by optical tweezers. Measuring the
distance of the particle from the wall the distribution of work
performed during one cycle of the potential modulation was determined
(histogram in Fig.4 in \cite{BSHSB}). Since $V_0(x)=V_1(x)$ this case
is characterized by $\Delta F=0$. 

\begin{figure}[t]
\begin{center}
\includegraphics[width=0.45\textwidth]{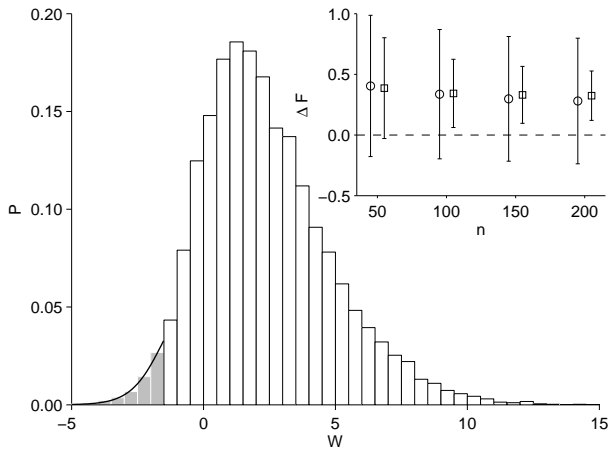}
\end{center}
\caption{\label{fig:hist3} Histogram of 16200 work values obtained
  experimentally in \cite{BSHSB} together with the asymptotic form 
  (full line) derived from (\ref{eq:LE3})-(\ref{eq:defD}). The inset
  shows the standard (circles) and the improved (squares) estimate for
  the free energy difference. The exact result is $\Delta F=0$ (dashed
  line).} 
\end{figure}

As discussed in \cite{BSHSB} the dynamics of the particle may be
approximately modeled by an overdamped Langevin equation. Due to the
vicinity of the wall the friction coefficient and the noise intensity
now depend on the state~$x$. Moreover, in order to retain the Gibbs
measure as stationary distribution of the stochastic process an
additional drift term has to be added \cite{LaLu}. Using It\=o
convention the resulting equation is \cite{Sascha} 
\begin{equation}
  \label{eq:LE3}
 \dot{x}=-D(x) V'(x,t) + D'(x)+\sqrt{2D(x)}\; \xi(t)\; ,
\end{equation}
with potential 
\begin{equation}
  \label{eq:defV3}
  V(x,t)=A e^{-\kappa (x-a)} + B(t)(x-a)
\end{equation}
and state dependent diffusion coefficient \cite{Brenner}
\begin{equation}
  \label{eq:defD}
  D(x)=\frac{D_0}{1+\frac{R}{x}}\; .
\end{equation}
The values for $D_0$, the radius $R$ of the particle, the
parameters $A, \kappa, a$ of $V(x,t)$, and the protocol function
$B(t)$ are taken from the experiment \cite{Blickle}. Instead of
(\ref{eq:defL}) we now have  
\begin{equation}
  \label{eq:defL3}
  L(x,\dotx,t)=\frac{1}{4D(x)}
        \Big(\dot{x}+D(x) V'(x,t)- D'(x)\Big)^2\; .
\end{equation}
The corresponding ELE 
\begin{align}\label{eq:ELE3}
  0=&\ddot{x}-\frac{D'}{2D}\dot{x}^2+(1-iq)D\dot{V}'\\\nonumber
    &+(D'-DV')(DV''-D''+\frac{1}{2}D'V'+\frac{D'^2}{2D})
\end{align}
can no longer be solved analytically but its numerical solution does
not pose any specific problems \cite{Sascha}. Solving (\ref{eq:ELE3})
for a wide range of $q$-values and using the solution in
(\ref{eq:defwork}) to establish the relation between $q$ and $W$ the
extremal action $\bar{S}$ can be determined. The calculation of the 
pre-exponential factor is now much more involved since the
differential equation for $Q(t)$ is more complicated and both $Q(t_1)$
and $R$ will depend on $W$ in a non-trivial way. We leave this
problem for further investigations and use for the present 
example only the asymptotic behaviour of $P(W)$ resulting from
(\ref{eq:asym1}). It is shown in Fig.~\ref{fig:hist3} together
with the histogram of experimental results. The inset gives again
a comparison of the estimates for $\Delta F$ determined analogously to
(\ref{eq:Jest}) and (\ref{eq:impest}) with $W^*=-3/2$. 

While the asymptotic form of $P(W)$ seems to be well captured the
improvement in the free energy estimates is less 
distinctive than in Fig.~\ref{fig:hist}. The reason may be that $P(W)$
decreases rather rapidly for small $W$ which makes the matching
between histogram and Asymptotics more difficult, in particular since
the pre-exponential factor is not available. On the other hand,
eq.~(\ref{eq:LE3}) is already an approximation to the experimental
situation and the asymptotic form of $P(W)$ derived from it may
therefore differ from the true one. 

\section{Conclusion}

We have shown that the method of optimal fluctuations 
allows to analytically characterize the asymptotic form of the work
distribution in driven Langevin systems. This information may be
combined with histograms of work values as obtained in experiments or
numerical simulations to improve the accuracy of free energy estimates
exploiting the Jarzynski equation. The method will work best in
situations where an overlap region in $W$-values exists which is 
sufficiently sampled by the histogram and at the same time well
described by the asymptotic behaviour. 

Our method builds on a saddle-point calculation of a functional
integral over stochastic trajectories constrained to a specific value
of the performed work $W$. Although similar techniques have been used
in the context of non-equilibrium work and fluctuation theorems (see,
e.g., \cite{TaCo,BaCo,MiAd}) the application to constrained problems
aiming at the {\it asymptotic} behaviour of the work distribution is
to our knowledge new. It will be interesting to generalize the method
to higher-dimensional situations. 

\vspace{2ex}

{\bf Acknowledgments:} I would like to thank Sascha von Egan-Krieger
for the fruitful collaboration on the issues discussed in this
paper. I have also benefited from interesting discussions with
Daniel Grieser, Aljoscha Hahn, Peter Reimann and Holger Then. Thanks
are due to Valentin Blickle for sending us the detailed parameters
and relevant results of his experiments.

\appendix

\section{}

In this appendix we give some details on the calculation of the
Gaussian fluctuations around the saddle-point in the integral
(\ref{eq:PofW}). Using $\eps=t_1/N$, $t_j=\eps j$, 
$V_j=V(\bx(t_j),t_j)$, and similar for the derivatives of
$V(\bx(t),t)$ the time-sliced version of this integral reads
\begin{widetext}
\begin{equation}
  P(W)=\lim_{N\to\infty}\frac{\beta}{4\pi Z_0}
        (\frac{\beta}{4\pi\eps})^{\frac{N}{2}}
          \int\!\! dq\int \prod_{j=0}^{N} dx_j 
              \; e^{-\beta S_N(\{x_j\},q)}
\end{equation}
with the discretized action defined by 
\begin{equation}\label{eq:SN}
  S_N(\{x_j\},q)=
        V_0+\frac{\eps}{2}\sum_{j=0}^{N-1}\Big[\frac{1}{2}
            (\frac{x_{j+1}-x_j}{\eps}+V'_j)^2+iq\dV_j\Big]
           -\frac{i}{2}qW \, .
\end{equation}
\end{widetext}
Denoting the saddle-point values of $x_j$ and $q$ by an overbar, 
using $\bS=S_N(\{\bx_j\},\bq)$ and expanding the exponent to second
order in $x_j-\bx_j$ and $q-\bq$ we find 
\begin{equation}
  P(W)= \lim_{N\to\infty}\frac{\eps}{Z_0}
    \frac{e^{-\bS}}{\sqrt{\det M}}\,(1+{\cal O}(1/\beta))
\end{equation}
where the symmetric matrix $M$ is given by 
\begin{align}\nonumber
  M_{kl}&=2\eps\,\frac{\pa^2 S_N}{\pa x_k \pa x_l}=:A_{kl} 
         \quad\text{for}\quad k,l=0,...,N\\\nonumber
  M_{k N+1}&=2\eps\,\frac{\pa^2 S_N}{\pa x_k \pa q}=
     i\eps^2\dot{\bar{V}}'_k \quad\text{for}\quad k=0,...,N\\\nonumber
  M_{N+1 N+1}&=2\eps\,\frac{\pa^2 S_N}{\pa q\pa q}=0 \; .
\end{align}
Here $A_{kl}$ is a tridiagonal fluctuation matrix of the usual form
\cite{Montroll,GeYa,ChDe}. Its determinant can be obtained from a
recursion relation which for $\eps\to 0$ turns into a differential
equation. Analogous to the standard Gelfand-Yaglom procedure we find
$\det A=Q(t_1)$ where $Q(t)$ is the solution of the initial value
problem (\ref{eq:resQ}).  

In order to reduce the calculation of $\det M$ to that of $\det A$ we
multiply the first $N+1$ rows of $M$ by 
$-i\eps^2(\dot{\bar{V}}')^T A^{-1}$ and add this to the last
row. The resulting matrix has then in the last row all zeros except
for the last entry which reads 
\begin{equation}
  R_N:=\eps^4\sum_{k,l}\dot{\bar{V}}'_k\, (A^{-1})_{kl}
        \,\dot{\bar{V}}'_l \, .
\end{equation}
Consequently $\det M= R_N \det A$. This result is in fact quite
intuitive. Assume for simplicity that the constraint $W[x(\cdot)]=W$
is orthogonal to one eigenvector $\mathbf{e}_n$ of $A$ with
eigenvalue $\lambda_n$. Then $\{\dot{\bar{V}}'_k\}$ which is the
gradient of the constraint is parallel to $\mathbf{e}_n$ and $R_N$ is
proportional to $1/\lambda_n$. It hence cancels exactly that
eigenvalue of the unconstrained fluctuation matrix $A$ describing 
fluctuations perpendicular to the constraint which are forbidden. 

Using 
\begin{equation}
  \lim_{N\to\infty} R_N=\eps^2 R
\end{equation}
with $R$ given by (\ref{eq:defR}) we finally end up with
(\ref{eq:asym}).

\end{document}